\documentclass[prl,twocolumn,epsf,psfig]{revtex4}
\usepackage{graphicx}
\DeclareGraphicsExtensions{.pdf,.png,.gif,.jpg}

\begin{document}

\title{Density profiles and density oscillations of an interacting three-component normal Fermi gas}

\author{Theja N. De Silva}
\affiliation{Department of Physics, Applied Physics and Astronomy,
The State University of New York at Binghamton, Binghamton, New York
13902, USA.}
\begin{abstract}
We use a semiclassical approximation to investigate density
variations and dipole oscillations of an interacting three-component
normal Fermi gas in a harmonic trap. We consider both attractive and
repulsive interactions between different pairs of fermions and study
the effect of population imbalance on densities. We find that the
density profiles significantly deviate from those of non-interacting
profiles and extremely sensitive to interactions and population
imbalance. Unlike for a two-component Fermi system, we find density
imbalance even for balanced populations. For some range of
parameters, one component completely repels from the trap center
giving rise a donut shape density profile. Further, we find that the
in-phase dipole oscillation frequency is consistent with Kohn's
theorem and other two dipole mode frequencies are strongly effected
by the interactions and the number of atoms in the harmonic trap.
\end{abstract}

\maketitle

\section{I. Introduction}
The recent experimental progress of trapping and cooling of atomic
gases leads to an opportunity for a detail study of exciting many
body physics~\cite{ex}. The key to this exciting opportunity is that
the high experimental controllability of these systems. For a dilute
atomic mixture, the quantum degenerate regime can be achieved by
cooling the system to ultra-cold temperatures. In these ultra-cold
atomic systems, one of the fascinating control parameters has been
the two-body scattering length of atoms. Tuning the scattering
length, simply by applying a homogenous magnetic field (Feshbach
resonance~\cite{fb}), the interactions can be controlled
dramatically. The fundamental differences between fermions and
bosons take place in the quantum degenerate regime. For the case of
two-component Fermi systems, if the temperature is low enough,
weakly bound molecules could be formed at positive scattering
lengths and these molecules can undergo Bose-Einstein condensation.
By sweeping the magnetic field, the scattering length can be further
increased to a divergency and can be changed in sign. In this
strongly interacting limit, the bosonic molecules continuously
transform into Bardeen, Cooper and Schrieffer (BCS) superfluid
pairs. The physics of this so-called BCS-BEC crossover region is
relevant to the systems like high $T_C$ superconductors, superfluid
$^3$He, and neutron stars. The recent realization of three-component
degenerate Fermi gases has opened a new research frontier in
ultra-cold atoms. These three-component systems have relevance to
high density quark matter, neutron stars, and universal few body
physics. Therefore, superfluid phases and phase separation in a
three-component Fermi gas can be treated as an analogous to color
superconductivity and baryon formation in quantum chromo dynamics.

\begin{figure}
\includegraphics[width=\columnwidth]{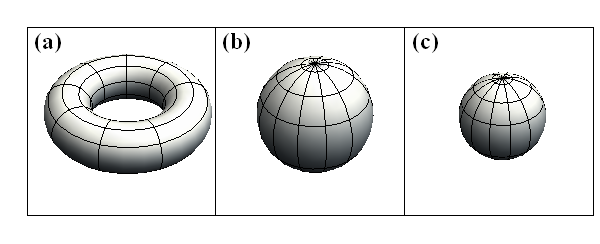}
\caption{Schematic atomic cloud shapes of a population imbalanced
three-component Fermi gas in a spherically symmetric harmonic trap.
Depending on the interaction between different pairs of fermions and
the population imbalance, one component can completely repel from
the trap center as shown in panel (a). The other two components form
a spherical shape as shown in panels (b) and (c). See
FIG.~\ref{density2} for quantitative details. For a population
balance mixture, inner radius of the donut shape cloud is zero and
we find density imbalance through out the harmonic trap.}
\label{shapes}
\end{figure}

In a recent experiment by Bartenstein \emph{et al}~\cite{grimm},
using radio frequency spectroscopic data, molecular binding
energies, scattering lengths and the Feshbach resonance positions of
the lowest three channels of $^6$Li atoms have been determined. As
there are three broad s-wave Feshbach resonances for the three
lowest hyperfine states of $^6$Li system, one can prepare the system
at various atomic interactions between different pairs. In another
recent experimental development by Ottenstein et al~\cite{jochim}, a
three-component Fermi gas has been cooled down to the quantum
degenerate regime. In this experiment, the three lowest hyperfine
states of $^6$Li atoms are prepared at a temperature $T = 0.37 T_F$,
where $T_F$ is the Fermi temperature. During the experiment, the
scattering lengths for the respective three Feshbach channels are
varied by sweeping a magnetic field and collisional stability of the
lowest three channels of $^6$Li atoms has been studied. The most
recent experiment~\cite{ohara} reports the first measurements of
three-body loss coefficients for a three-component Fermi gas at
temperature $T = 0.59 T_F$. By further lowering the temperature,
paring of three components can be activated. This promising
experimental observations of three-component superfluidity is yet to
be realized in near future. Motivated by the realization of these
normal degenerate three-component Fermi systems~\cite{grimm, jochim,
ohara, jochimA}, we study the ground state normal density profiles
and dipole density oscillations of a three-component Fermi gas
trapped in a harmonic potential at zero temperature. We use a
semiclassical theory known as Thomas Fermi functional approach to
investigate the normal state density profiles and dipole density
oscillations. In general, at high enough temperatures, the particles
are classical so that the semiclassical approximation is valid. In
the present paper, we are considering a large number of Fermi atoms.
These atoms fill every energy levels up to the Fermi energy $E_F$
and $E_F$ is larger than the ground state energy. Therefore, a
three-component normal Fermi gas can be well described in the
semiclassical approximation even at very low temperatures provided
that the number of atoms present in the system is large.

For an equal population system, we find that the density imbalanced
cloud form a three-shell structure. Depending on the interactions
between pairs, density of the one component shows non -monotonic
behavior as the cloud extends from center to the edge of the
harmonic trap. For an unequal population system, we find very
similar behavior as the population balanced system, however, one
component completely repels from the trap center. This repulsion
depends on both interaction and the population imbalance. The
generic shapes of the atomic clouds in a harmonic trapping potential
for this case are schematically shown in FIG.~\ref{shapes}. In the
presence of harmonic trapping potential, we derive expressions for
in-phase and and out-of-phase dipole oscillation frequencies as
functions of interactions and atom numbers in different spin states.
we find that the in-phase oscillation frequency is independent of
the interactions and the atom numbers while the out-of-phase
oscillation frequencies strongly depend on them.

The paper is organized as follows. In the following section, we
present the basic formalism required for the densities and dipole
oscillation frequencies. In section III, we derive the equations for
Thomas Fermi density profiles and stability condition and then
present our self consistent solutions based phase diagram and the
density profiles. In section IV, we present dipole oscillation
frequencies as a function of interactions and atom numbers followed
by the discussions and summary in section V.

\section{II. Basic Formalism}
We assume that the three-component Fermi gas is trapped in a
harmonic potential given by $V(r)=(\omega_\perp r_1^2+ \omega_\perp
r_2^2+\omega_z r_3^2)/(2m)$, where $\omega_\perp$, $\omega_z$, and
$m$ are transverse and axial trapping frequencies and mass of an
atom respectively. Here $r_i$ is the $i$th cartesian component of
the position vector $\vec{r}$. Let us denote the density of the
$\sigma$ component is $n_{\sigma}$ and the interaction strength
between $\sigma$ and $\nu$ components is $U_{\sigma \nu}$.

As we are considering a dilute atomic system at zero temperature,
and the interactions are short-range in nature, s-wave scattering
channel is dominated over the other scattering channels. Therefore,
we can neglect the higher wave scatterings and the interaction can
be specified by a single parameter $a_{\sigma \nu}$, which is the
s-wave scattering length between component $\sigma$ and $\nu$. The
functional form  of the interaction is $U_{\sigma \nu}= 4\pi
\hbar^2a_{\sigma \nu}/m$. In order to derive the ground state
densities of the normal state at zero temperature, we write the
grand canonical energy as a functional of densities, $E_0 = E_g -
\sum_{\sigma =1}^3\mu_\sigma N_\sigma$. Here the number of atoms in
each spin state $N_\sigma = \int d^3\vec{r}n_\sigma$ and the total
energy $E_g = \int \epsilon d^3\vec{r}$, where $\epsilon$ is given
by

\begin{eqnarray}\label{model1}
\epsilon = \sum_{\sigma=1}^3\biggr\{\frac{\hbar^2A}{2m}n_\sigma ^{5/3}+V(r)n_\sigma \biggr\} + \nonumber \\
\frac{4\pi\hbar^2}{m}\biggr \{a_{23}n_2n_3 + a_{13}n_1n_3 +
a_{12}n_1n_2\biggr\},
\end{eqnarray}

\noindent and $A \equiv (3/5) (6\pi^2)^{2/3}$. Notice we introduced
three Lagrange multipliers $\mu_\sigma$ (chemical potentials) to
constrain the number of atoms in each $\sigma$ state. In section III
presented below, we derive Thomas-Fermi equations for the densities
from the variation of the total energy.

In the presence of harmonic trapping potential, the lowest lying
collective excitations are the dipole or the center of mass
oscillations.  By generalizing the scaling method used in
Ref.~\cite{2comptf3}, we derive expressions for the dipole
oscillation frequencies for a three-component system. We assume that
the atomic cloud is displaced linearly along $r_3$ direction. The
time varying density of the $\sigma$ component is parameterized by
the collective coordinate $\eta_\sigma (t)$ as $n_\sigma(r,t) =
n^{(0)}_\sigma(r_1,r_2,r_3-\eta_\sigma)$, where $n^{(0)}_\sigma (r)$
is the ground state equilibrium density distribution of the $\sigma$
component. Substituting $n_\sigma(r,t)$ into total energy
expression, the variation of the total energy up to the quadratic
order in $\mathbf{\eta}$ is given by

\begin{eqnarray}\label{ev}
\Delta E = E-E_0 \approx -\frac{1}{2}m\omega^2_z \sum_\sigma
\eta_\sigma^2 N_\sigma \nonumber \\
+ A_{12}(2\eta_1\eta_2-\eta^2_1-\eta_2^2) \nonumber \\
+ A_{13}(2\eta_1\eta_3-\eta^2_1-\eta_3^2) \nonumber \\
+ A_{23}(2\eta_2\eta_3-\eta^2_2-\eta_3^2)
\end{eqnarray}

\noindent where

\begin{eqnarray}\label{Aex}
A_{\sigma \nu} = \frac{U_{\sigma \nu}}{2}\int d^3\vec{r}
\frac{\partial n^{(0)}_\sigma}{\partial r_3}\frac{\partial
n^{(0)}_\nu}{\partial r_3}.
\end{eqnarray}

\noindent The dipole oscillation frequencies ($\omega_d$) are
determined by the eigenvalues of the classical equations of motion
for $\mathbf{\eta}$. The classical equation of motion in the matrix
form is given as,

\begin{widetext}
\begin{eqnarray}\label{mat}
\left(
\begin{array}{ccc}
m(\omega_z^2-\omega_d^2)N_1-2A_{12}-2A_{13} & 2A_{12} & 2A_{13} \\
2A_{12} & m(\omega_z^2-\omega_d^2)N_2-2A_{12}-2A_{23} & 2A_{23} \\
2A_{13} & 2A_{23} &
m(\omega_z^2-\omega_d^2)N_3-2A_{13}-2A_{23}\\
\end{array}
\right) \left( \begin{array}{c}
 \eta_1 \\
 \eta_2 \\
\eta_3 \\
 \end{array}
\right) = 0.
\end{eqnarray}
\end{widetext}

\section{III. Thomas-Fermi equations for densities}

Before we derive the Thomas-Fermi equations, we introduce three
dimensionless variables; $\rho_{\sigma} \equiv n_{\sigma}\zeta^3$,
$\tilde{r}^2 \equiv \tilde{r}_1^2+\tilde{r}_2^2+\tilde{r}_3^2$ and
$\tilde{a}_{\sigma \nu} \equiv (4 \pi)^2a_{\sigma \nu}/\zeta$ where
$\zeta = \sqrt{\hbar/m\omega}$ is the effective oscillator length
with $\omega = (\omega_\perp^2 \omega_z)^{1/3}$. The dimensionless
cartesian component of the vector $\vec{r}$ are defined as
$\tilde{r}_i \equiv [2\pi m/(\hbar \omega)\omega_i^2]^{1/2}r_i$. In
terms of dimensionless variables, now the trapping potential has a
symmetric form given by $V(r) = \hbar\omega \tilde{r}^2/(4\pi)$. In
terms of dimensionless parameters, the number equation and the total
energy are given by

\begin{eqnarray}\label{number}
N_\sigma = 4 \pi\int \tilde{r}^2 \rho_\sigma d\tilde{r},
\end{eqnarray}

\noindent and

\begin{eqnarray}\label{model2}
\frac{E}{\hbar\omega} = \int \tilde{r}^2 d\tilde{r}\biggr \{ \sum_{\sigma=1}^3\biggr(2 \pi A\rho_\sigma^{5/3} + \tilde{r}^2\rho_\sigma -\tilde{\mu}_\sigma \rho_\sigma\biggr) \nonumber \\
+ \tilde{a}_{23}\rho_2\rho_3 + \tilde{a}_{13}\rho_1\rho_3 +
\tilde{a}_{12}\rho_1\rho_2\biggr \}.
\end{eqnarray}

\noindent where the dimensionless chemical potential
$\tilde{\mu}_\sigma = 4 \pi \mu_\sigma/(\hbar \omega)$. The
Thomas-Fermi equations for the densities $\rho_\sigma$ are derived
from the variation of the total energy; $\partial E/\partial
\rho_{\sigma} = 0$.

\begin{eqnarray}\label{tfeq}
\frac{10 \pi A \rho_1^{\frac{2}{3}}}{3} + \tilde{a}_{13}\rho_3 + \tilde{a}_{12}\rho_2 = \tilde{\mu}_1 - \tilde{r}^2 \nonumber \\
\frac{10 \pi A \rho_2^{\frac{2}{3}}}{3}  + \tilde{a}_{23}\rho_3 + \tilde{a}_{12}\rho_1 = \tilde{\mu}_2 - \tilde{r}^2 \nonumber \\
\frac{10 \pi A \rho_3^{\frac{2}{3}}}{3}  + \tilde{a}_{23}\rho_2 +
\tilde{a}_{13}\rho_1 = \tilde{\mu}_3 - \tilde{r}^2
\end{eqnarray}

\noindent For the spatial density variations, these equations must
be solved with the stability condition. The stability condition can
be derived from the determinant of second order variation of the
energy functional, $|\partial^2E /\partial \rho_\sigma \partial
\rho_\nu| \geq 0$, where $\partial^2E /\partial \rho_\sigma \partial
\rho_\nu$ is a $3 \times 3$ matrix. In terms of dimensionless
variables, this stability condition reads,

\begin{eqnarray}\label{scond}
\frac{2}{3}\biggr(\frac{10\pi A}{3}\biggr)
(\tilde{a}_{23}^2\rho_2^{\frac{1}{3}}\rho_3^{\frac{1}{3}} + \tilde{a}_{13}^2\rho_1^{\frac{1}{3}}\rho_3^{\frac{1}{3}} + \tilde{a}_{12}^2\rho_1^{\frac{1}{3}}\rho_2^{\frac{1}{3}}) \nonumber \\
- 2 \tilde{a}_{12}\tilde{a}_{23}\tilde{a}_{13}
\rho_1^{\frac{1}{3}}\rho_2^{\frac{1}{3}}\rho_3^{\frac{1}{3}} \leq
\biggr(\frac{2}{3}\frac{10\pi A}{3}\biggr)^3
\end{eqnarray}

Similar set of equations for an repulsively interacting
two-component normal gas in three dimensions are derived in
Refs.~\cite{2comptf1, 2comptf2, 2comptf3}. For a two-component
normal gas in two dimensions, the Thomas Fermi equations are derived
in Ref.~\cite{2comptf2D}. We have investigated the stability
condition given in Eq. (\ref{scond}). Unlike two-component Fermi
gases, three-component Fermi mixtures are stable for large densities
and interactions~\cite{jochim}. For a dilute system, we do not
expect three body collisions to take place so that the system will
be stable for a wide range of interactions.

\begin{figure}
\includegraphics[width=\columnwidth]{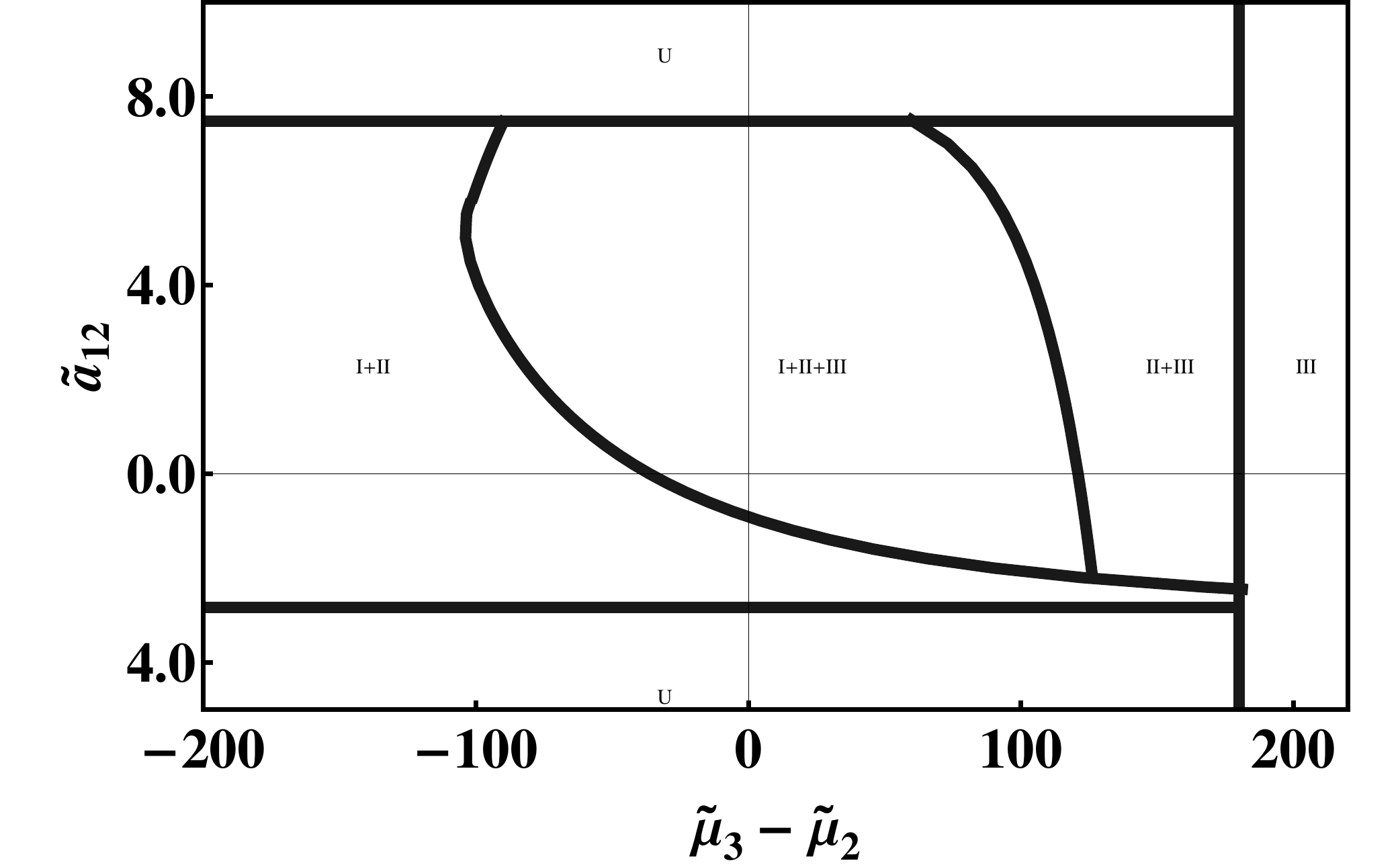}
\caption{Phase diagram for fixed $\mu_1 = (4\pi/\hbar \omega) 190 $,
$\mu_2 = (4\pi/\hbar \omega) 178 $, $\tilde{a}_{23} = 2.5$, and
$\tilde{a}_{13} = 3.6$. The labels I, II, and III are the phases
with component one, two, and three respectively. Label U denotes the
unstable region.} \label{pd1}
\end{figure}

Before we investigate the density variations, let us construct the
phase diagram for a homogenous system. We consider a set of
representative parameters by fixing the dimensionless chemical
potentials of components one and two as $\tilde{\mu}_1 = 190$ and
$\tilde{\mu}_2 = 178$ (these values are corresponding to the central
chemical potentials of the system with particle numbers in the order
of $10^5$). Further, we chose the interactions to be $\tilde{a}_{23}
= 2.5$ and $\tilde{a}_{13} = 3.6$. Then by solving the
equations~(\ref{tfeq}) and (\ref{scond}), the phase diagram is
constructed in $\tilde{a}_{12}$ - $(\tilde{\mu}_3-\tilde{\mu}_2)$
parameter space as shown in FIG.~\ref{pd1}. When there is no
interaction between component one and two ($a_{12} =0$), the phase
boundaries can be easily evaluated analytically. Let us define the
possible phases by the notation $i =$ I, II, and III, where $i$ is
the phase with component $i$. The mixed phases are defined as sum of
$i$'s (For example, I+II is the phase where both component one and
two exist). The phase boundary between I+II and I+II+III phases is
given by $\tilde{\mu}_3 =
\beta_{23}\tilde{\mu}_2^{3/2}+\beta_{13}\tilde{\mu}_1^{3/2}$, where
$\beta_{\sigma \nu} = 3 \tilde{a}_{\sigma \nu}/(10 \pi A)$. The
boundary between I+II+III and II+III phases is given by
$\tilde{\mu}_3 = (\tilde{\mu}_1/\beta_{13})^{2/3} +
\sqrt{\tilde{\mu}_2 -
\tilde{\mu}_1\beta_{23}/\beta_{13}}(\tilde{\mu}_2
\beta_{23}-\tilde{\mu}_1\beta^2_{23}/\beta_{13})$. The phase
boundary between II + III and III phases is given by $\tilde{\mu}_3
= (\tilde{\mu}_2/\beta_{23})^{2/3}$. Even though, we have restricted
ourselves to the repulsive interactions between two-three and
one-three pairs, one can consider attractive interactions between
these pairs and construct the phase diagram in similar manner.

\begin{figure*}
\includegraphics[width=\textwidth,clip]{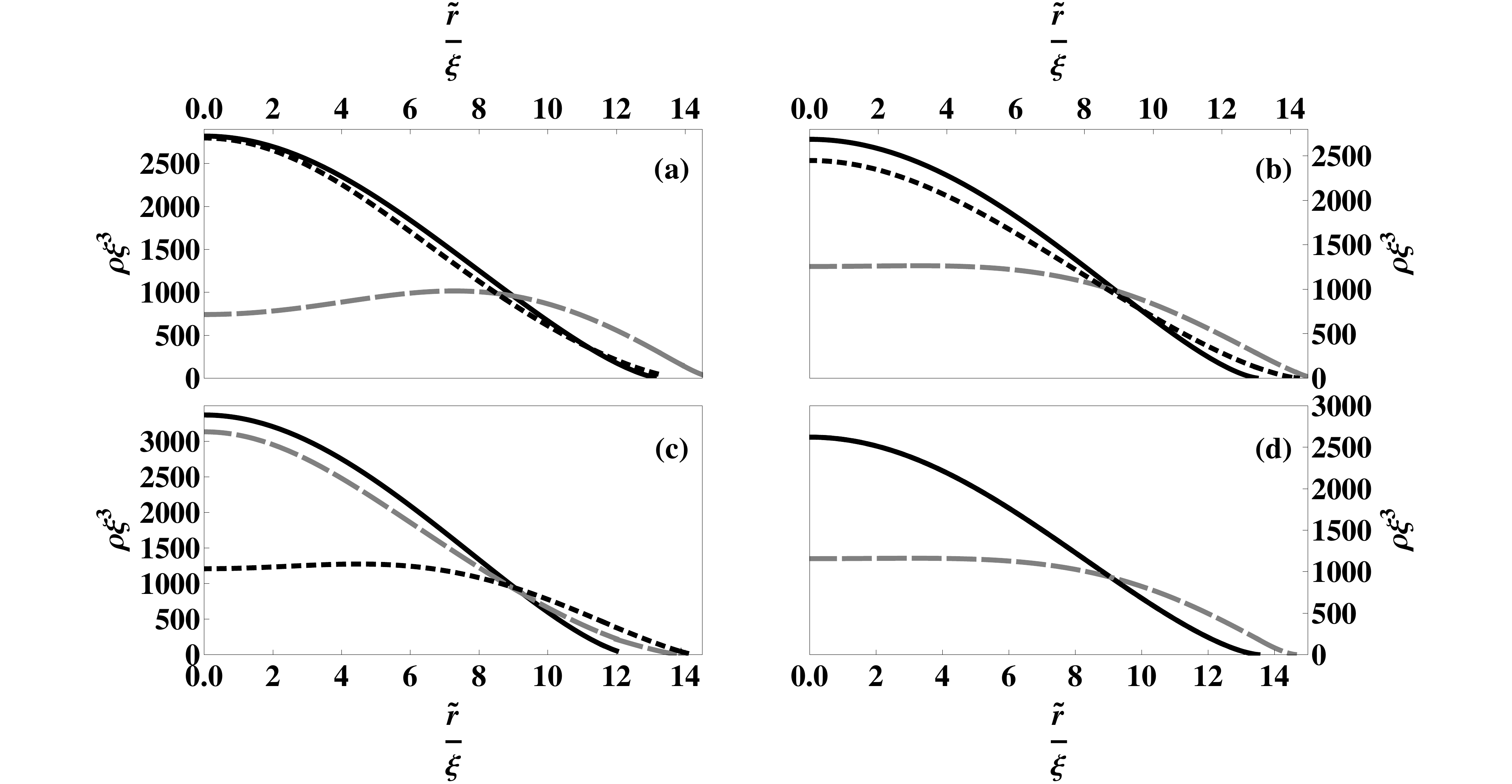}
\caption{Density profiles of the three Fermi components for various
interactions between pairs. We fixed the number of atoms in each
spin state to be $N_\sigma \simeq 4.5 \times 10^5$. The
dimensionless scattering lengths between pairs are: (a)
$\tilde{a}_{12} = 1$, $\tilde{a}_{13} = -1$, and $\tilde{a}_{23} =
4$, (b) $\tilde{a}_{12} = 0.1$, $\tilde{a}_{13} = -1$, and
$\tilde{a}_{23} = 4$, (c) $\tilde{a}_{12} = -2$, $\tilde{a}_{13} =
-1$, and $\tilde{b}_{23} = 4$, (d) $\tilde{a}_{12} = 1$,
$\tilde{a}_{13} = 2$, and $\tilde{a}_{23} = 2$. Black line:
component 1, dashed gray line: component 2, and dotted line:
component 3.} \label{density1}
\end{figure*}

For density variations, we simultaneously solve equations
(\ref{number}), (\ref{tfeq}), and (\ref{scond}) for given atom
numbers in each spin states and given interactions between different
pairs. In figure~\ref{density1}, we plot the density profiles for
various interaction strengths for an atomic mixture with equal
numbers in each three spin states. As seen in the figure, all three
components exist in the trap center. However, the density imbalance
remains finite through out the trap. The most pronouncing feature is
the non-monotonic density variations of one component when both
repulsive and attractive interactions are present. As a result of
the spatially inhomogeneous trap, the balanced mixture of atomic
cloud forms a 3-shell structure similar to the case of population
imbalanced two-component gases. While the inner most shell contains
all three components, the outer most shell contains only a single
component. Even though, we consider a population \emph{balanced}
mixture in FIG.~\ref{density1}, we find density \emph{imbalanced}
throughout the trap, similar to a case of population imbalanced
two-component system. It should be noted that the density imbalance
in three-component system here is not due to the population
imbalance but due to the different interactions between different
pairs.

\begin{figure*}
\includegraphics[width=\textwidth,clip]{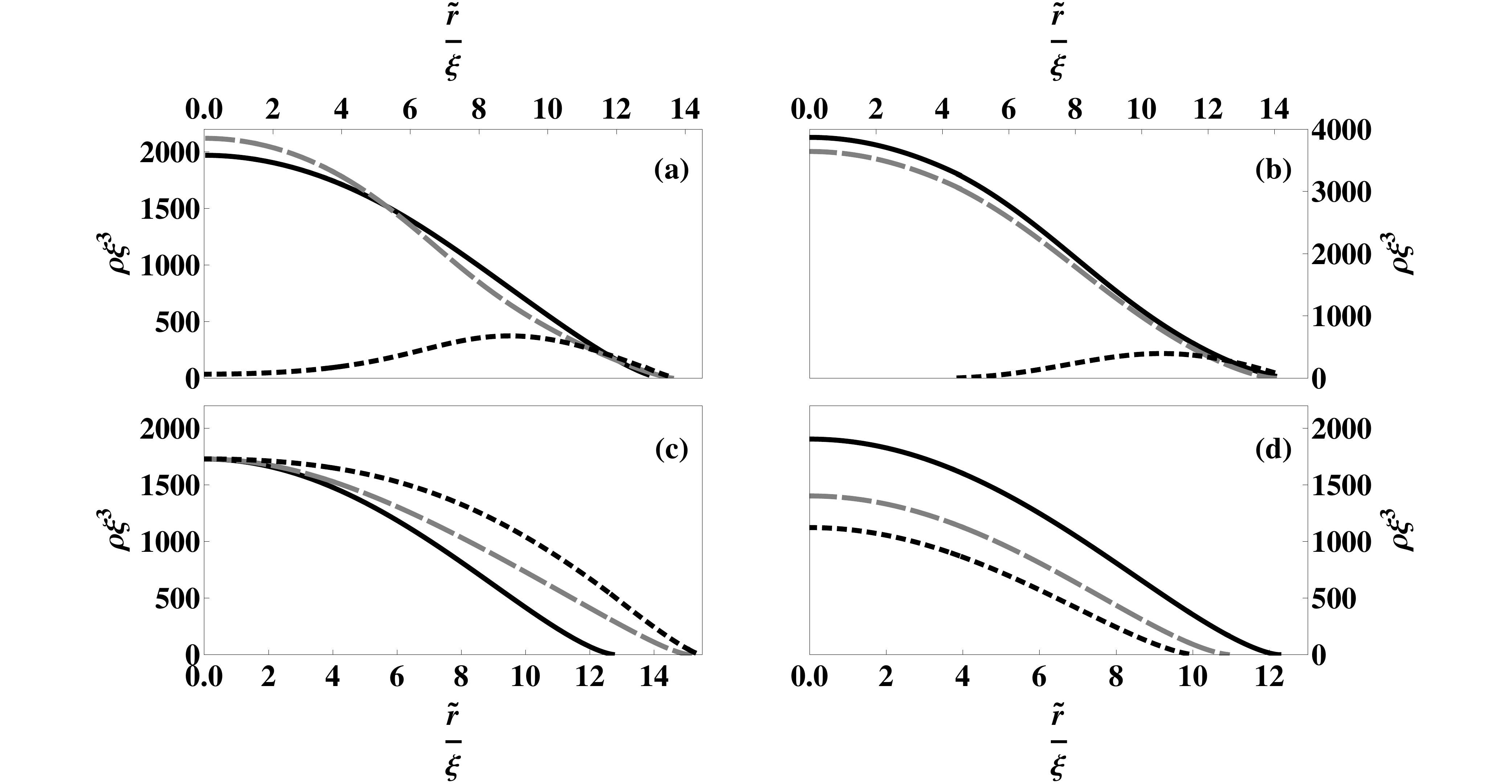}
\caption{Density profiles of the three Fermi components for various
interactions between pairs and different numbers. The number of
atoms and the dimensionless scattering lengths between pairs are:
(a) $\tilde{a}_{12} = 1$, $\tilde{a}_{13} = 2$, and $\tilde{a}_{23}
= 6.6$, $N_1 \simeq 3.0 \times 10^5$, $N_2 \simeq 3.0 \times 10^5$,
and $N_3 \simeq 1.6 \times 10^5$ (b) $\tilde{a}_{12} = -1$,
$\tilde{a}_{13} = 3.6$, and $\tilde{a}_{23} = 2.5$, $N_1 \simeq 6.7
\times 10^5$, $N_2 \simeq 6.0 \times 10^5$, and $N_3 \simeq 1.9
\times 10^5$ (c) $\tilde{a}_{12} = 0.462$, $\tilde{a}_{13} = 1.04$,
and $\tilde{a}_{23} = 4.51$, $N_1 \simeq 2.8 \times 10^5$, $N_2
\simeq 4.7 \times 10^5$, and $N_3 \simeq 6.7 \times 10^5$ (d)
$\tilde{a}_{12} = -0.1$, $\tilde{a}_{13} = -0.2$, and
$\tilde{a}_{23} = -0.3$ $N_1 \simeq 2.7 \times 10^5$, $N_2 \simeq
1.4 \times 10^5$, and $N_3 \simeq 8.8 \times 10^4$. Black line:
component 1, dashed gray line: component 2, and dotted line:
component 3.} \label{density2}
\end{figure*}

The density profiles of an atomic mixture with unequal populations
are shown in FIG.~\ref{density2}. We find very similar density
profiles as the case of equal population, however, depending on the
number of atoms in each spin states and the interaction between
different pairs, one component can completely repel from the trap
center giving rise a donut shape density profile. This donut shape
density profile is an unique feature of a three-component normal gas
in the presence of population imbalance where as this feature is
absent in a two-component gas.

For the data in FIG~\ref{density1} and FIG.~\ref{density2}, we
choose interaction parameters unsystematically to show the different
behavior of density profiles. Even though the density profiles are
shown as a function of scaled parameter $\tilde{r}$, the generic
behavior in real space must be the same. This can be easily
understood if we consider the isotropic case where $\omega_\perp =
\omega_z \equiv \omega$. For this case, the atomic cloud is
isotropic and the density variations can be presented as a function
of $r$, just like in FIG~\ref{density1} and FIG.~\ref{density2}.

In experiments, real-space density distributions of each components
can be obtained by using \emph{in situ} imaging. After trapping and
cooling, the system can be prepared at any desirable population. As
experimental time scales are smaller than the spin relaxation time
of the atoms, fixed spin population can be maintained throughout the
experiment. Then by sweeping an uniform magnetic field, scattering
lengths between different pairs can be set at different values. As
some three-component atomic systems (for example, lowest hyperfine
spin states of $^6$Li~\cite{jochim}) possess three broad s-wave
Feshbach resonances, by tuning  scattering lengths, the interactions
between different pairs can be varied. However, in a current
experimental setups, interaction strengths, (i.e, the scattering
lengths between different pairs) cannot be controlled independently
whereas the number of atoms in each spin states can be varied
independently.

Density profiles of an repulsively interacting two-component atomic
gas in a harmonic trap are discussed in Ref.~\cite{2comptf1,
2comptf2, 2comptf2D, 2compsc} in the context of collective
ferromagnetism. While the authors in these references discuss
spontaneous magnetization and phase separations of two hyperfine
spin components, Duine \emph{et al}~\cite{macd} discuss the nature
of the ferromagnetic phase transition in the mean field description
and beyond mean filed theory of a homogenous system. The study in
Ref.~\cite{2comptf2} is similar to the study in
Refs.~\cite{2comptf1, 2compsc}, however, authors in
Ref.~\cite{2comptf2} go beyond mean field theory and study the
effect of trap anisotropy on magnetization. Except
Ref.~\cite{2compsc}, all above studies are restricted to
two-component atomic gases. Salasnich \emph{at el}~\cite{2compsc}
investigate the density profiles of a three-component gas, however,
the authors have restricted themselves to an equal and repulsive
interaction between different pairs of fermions. As we have relaxed
their constraints in the present study allowing to have both
repulsive and attractive, as well as different interactions, we find
qualitatively and quantitatively very different results.

\section{IV. Dipole oscillations}

By solving for the eigenvalues of Eq.~\ref{mat}, we find the three
dipole oscillation frequencies, $\omega_0 = \omega_z$ and
$\omega_\pm = \sqrt{\omega_z^2-C \pm \sqrt{B}}$, where

\begin{eqnarray}\label{c}
C = \frac{1}{2}\sum_{\sigma \ne \nu}K_{\sigma \nu} (N_\sigma +
N_\nu)
\end{eqnarray}

\noindent and

\begin{eqnarray}\label{b}
B = \frac{1}{2}\sum_{\sigma \ne \nu}K_{\sigma \nu}^2 (N_\sigma +
N_\nu) + \nonumber \\
\frac{1}{2}\sum_{\sigma \ne \nu \ne \mu} K_{\sigma \nu} K_{\sigma
\mu} (N_\nu N_\mu - N_\sigma N_\mu - N_\sigma N_\nu - N^2_\sigma).
\end{eqnarray}

\noindent The parameter $K_{\sigma\nu} = A_{\sigma\nu}/(m N_\sigma
N_\nu)$. The eigenvalue $\omega_0$ corresponding to the eigenvector
$\eta_1 = \eta_2 = \eta_3$ and represents the in-phase oscillation
frequency of the system. This mode follows the Kohn's theorem and it
is independent of interactions and the atomic population. The other
two eigenvectors are $\eta_1/N_1 = \eta_2/N_2 = -\eta_3/N_3$ and
$\eta_1/N_1 = -\eta_2/N_2 = \eta_3/N_3$. These two modes are
corresponding to the oscillations where two components are in
in-phase while the other one is in out-of-phase oscillations.

In terms of dimensionless parameters, $K_{\sigma \nu}$ is given by

\begin{eqnarray}\label{kv}
K_{\sigma \nu} = \frac{3 \pi \tilde{a}_{\sigma
\nu}\omega_z^2}{N_\sigma N_\nu}\int \tilde{r}^2 d\tilde{r}
\frac{\partial \rho^{(0)}_\sigma}{\partial \tilde{r}}\frac{\partial
\rho^{(0)}_\nu}{\partial \tilde{r}}.
\end{eqnarray}

\begin{figure}
\includegraphics[width=\columnwidth]{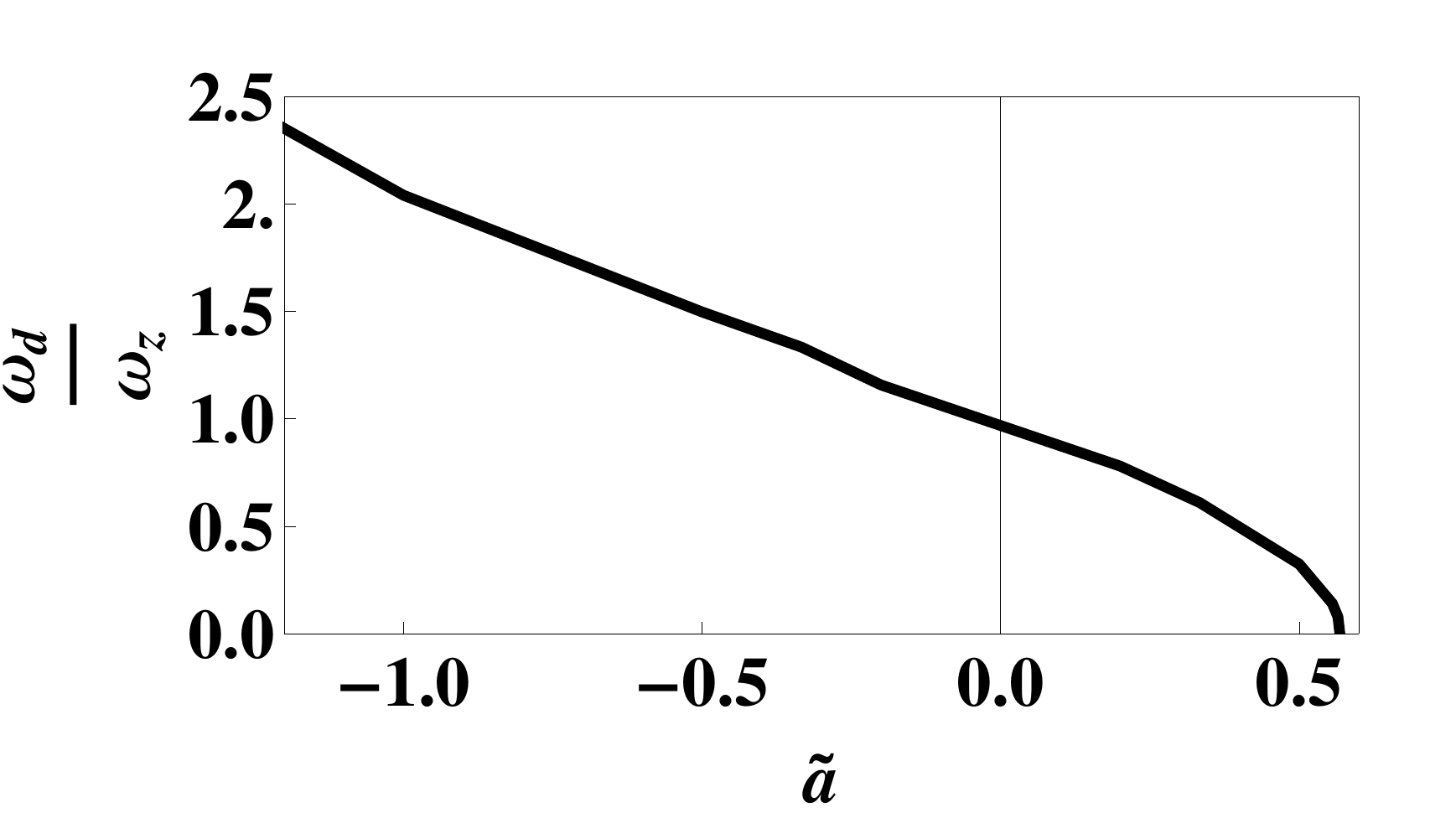}
\caption{Out-of-phase dipole oscillation frequency as a function of
interaction ($\tilde{a}$) for a symmetric system with $N = 6.5
\times 10^6$ number of atoms in each spin state. We set the
interaction between different pairs of fermions to be equal.}
\label{dpm}
\end{figure}

\noindent In order to see the qualitative behavior of the
out-of-phase oscillations, we restrict ourself to the equal
population and equal interactions. For the case of equal population
$N_1 = N_2 = N_3 \equiv N$ and equal interactions $\tilde{a}_{12} =
\tilde{a}_{13} = \tilde{a}_{23} \equiv \tilde{a}$, the out-of-phase
modes are degenerate and given by

\begin{eqnarray}\label{dmodes}
\biggr(\frac{\omega_d}{\omega_z}\biggr)^2 = 1 - \frac{4 \pi
\tilde{a}}{3N}\int \tilde{r}^2 d\tilde{r} \biggr(\frac{\partial
\rho^{(0)}}{\partial \tilde{r}}\biggr)^2
\end{eqnarray}

\noindent where $\rho^{(0)} = \rho_1^{(0)} + \rho_2^{(0)} +
\rho_3^{(0)}$. This density oscillation frequency as a function of
interaction is shown in FIG.~\ref{dpm} for the case of $N = 6.5
\times 10^6$. For this case case, the size of the Fermi gas becomes
larger as the interaction increases, and then the density
oscillation frequency $\omega_d$ monotonically decreases.

As it has been suggested recently~\cite{taylor}, the out-of phase
dipole mode can be excited via two-photon Bragg scattering in
experiments. As the two-photon Bragg scattering spectrum is related
to the dynamic structure factor, measuring the density response
spectrum with available experimental techniques~\cite{bragg} allows
one to measure the out-of phase dipole mode frequencies discussed
here.

\section{V. Discussions and Summary}

Even though, we have restricted ourselves to the zero temperature in
this paper, we do not expect qualitative changes at finite
temperatures. Even with the attractive interaction between some
pairs, the system will be in normal state if the population
imbalance is large. In the presence of attractive interactions
between different pairs of fermions (and at low population
imbalances), it is possible to form s-wave superfluid phases which
require equal densities. Unlike a two-component Fermi gas, a
three-component Fermi gas can undergo competitive BCS pairing and
provide non-trivial order parameters. In the presence of a
superfluid phase, superfluid region will spatially phase separate
into a concentric shell surrounded by density imbalance normal
shells. Even though, the experimental observations of trionic bound
states and superfluid phases are yet to be observed in near future,
there are series of theoretical studies on three-component
superfluidity~\cite{3compsf1, 3compsf2, 3compsf3, 3compsf4,
3compsf5}.

Another promising perspective of a three-component Fermi system is
to observe universal three-body quantum physics called, Efimov
states. In the present work, we have neglected the appearance of an
infinite number of three-body bound states (Efimov states) and
two-body pairing between components. As there is no three-body
interactions, three-body recombination is forbidden in a dilute
two-component Fermi gas.

In summary, we use a density functional approach to investigate the
Thomas Fermi density profiles and dipole oscillation frequencies of
an interacting three-component normal Fermi gas at zero temperature.
We constructed a phase diagram for a homogenous system at selected
representative values of parameters to show that the system
possesses a rich phase diagram, even neglecting the possible pairing
between atom pairs. Even for an equal population atomic mixture, we
find density imbalance through out the harmonic trap due to the
different interaction between different pairs. For an unequal
population atomic mixture, we find that one of the component
completely repel from the trap center showing a donut shape atomic
cloud surrounded by two concentric spherical clouds. In both cases,
density profiles show dramatic deviations showing non-monotonic
density distributions as opposed to the distributions seen in a
two-component gas. Another qualitative difference from the
two-component gas is that the density imbalance even for a
population balanced normal gas. This density imbalance in the
presence of population balance is due to the different interactions
between different pairs. We also derive in-phase and out-of-phase
dipole oscillation frequencies to show how interaction and
population imbalance effect on these mode frequencies.

\end{document}